\documentclass[runningheads]{llncs}
\usepackage{graphicx}
\usepackage[nobiblatex]{xurl}
\usepackage{comment}
\usepackage{rotating}
\usepackage{adjustbox}

\usepackage{framed}
\usepackage{latexsym}
\usepackage{graphics}
\usepackage{xcolor}
\usepackage{booktabs}
\usepackage{bigstrut}
\usepackage{url}
\usepackage{multicol}
\usepackage{multirow}
\usepackage{verbatim}
\usepackage{tabularx}
\usepackage{color,soul}
\usepackage{subfigure}
\usepackage{hyperref}
\usepackage{booktabs}
\usepackage{bigstrut}
\usepackage{color,soul}
\usepackage{float}
\usepackage{amsmath}

 \graphicspath{ {./APA/} }

\usepackage{amsfonts}

\titlerunning{CADICA: a new dataset for coronary artery disease detection by using invasive coronary angiography}

\usepackage[
backend=biber,
style=numeric,
sorting=none
]{biblatex}
\begin{document}
\title{CADICA: a new dataset for coronary artery disease detection by using invasive coronary angiography}
\titlerunning{CADICA: a new dataset for coronary artery disease}
%
\author{Ariadna Jim\'enez-Partinen\inst{1,3} \and
Miguel A. Molina-Cabello\inst{1,3} \and
Karl Thurnhofer-Hemsi\inst{1,3,4} \and
Esteban J. Palomo\inst{1,3} \and
Jorge Rodríguez-Capitán\inst{2,3,4} \and
Ana I. Molina-Ramos\inst{2,3,4} \and
Manuel Jim\'enez-Navarro\inst{2,3,4,5}}
\authorrunning{Ariadna Jim\'enez-Partinen et al.}
%
\institute{Department of Computer Languages and Computer Science. University of M\'alaga, Bulevar Louis Pasteur, 35, M\'alaga, Spain, 29071 \and
Cardiology Deparment, Hospital Universitario Virgen de la Victoria, M\'alaga, 29010, Spain\\ \and
Instituto de Investigación Biomédica de Málaga y Plataforma en Nanomedicina-IBIMA Plataforma BIONAND, C/ Severo Ochoa, 35, Málaga TechPark, Campanillas, 29590, M\'alaga, Spain\\ \and
Centro de Investigaci\'on Biom\'edica en Red de Enfermedades Cardiovasculares (CIBERCV), Instituto de Salud Carlos III (ISCIII), Avenida Monforte de Lemos, 3-5. Pabell\'on 11. Planta 0, 28029, Madrid, Spain\\ \and
Facultad de Medicina, University of M\'alaga, Bulevar Louis Pasteur, 37, 29071,  M\'alaga, Spain\\}
\maketitle              
\begin{abstract}
Coronary artery disease (CAD) remains the leading cause of death globally and invasive coronary angiography (ICA) is considered the gold standard of anatomical imaging evaluation when CAD is suspected. However, risk evaluation based on ICA has several limitations, such as visual assessment of stenosis severity, which has significant interobserver variability. This motivates to development of a lesion classification system that can support specialists in their clinical procedures. Although deep learning classification methods are well-developed in other areas of medical imaging, ICA image classification is still at an early stage. One of the most important reasons is the lack of available and high-quality open-access datasets. In this paper, we reported a new annotated ICA images dataset, CADICA, to provide the research community with a comprehensive and rigorous dataset of coronary angiography consisting of a set of acquired patient videos and associated disease-related metadata. This dataset can be used by clinicians to train their skills in angiographic assessment of CAD severity, by computer scientists to create computer-aided diagnostic systems to help in such assessment, and to validate existing methods for CAD detection. In addition, baseline classification methods are proposed and analyzed, validating the functionality of CADICA with deep learning-based methods and giving the scientific community a starting point to improve CAD detection.

\end{abstract}

\section{Introduction} \label{sec:introduction}

Coronary artery disease (CAD) remains the leading cause of death globally \cite{murphy2021mortality, voudris_advances_2019}. Clinical presentations of CAD are currently categorized as either acute coronary syndromes or chronic coronary syndromes, and assessing patients with suspected CAD is a significant component of healthcare costs \cite{saraste_imaging_2019}. Invasive coronary angiography (ICA) is considered the gold standard of anatomical imaging evaluation when CAD is suspected \cite{knuuti_2019_2020, collet_2020_2021}. ICA acquisition is based on introducing radiocontrast through a catheter inserted by a percutaneous incision in the femoral or brachial artery, situated in the groin and the arm, respectively. The radiocontrast agent enhances the visibility of coronary arteries, with cardiac angiography equipment, X-ray-based, the state of the arteries is shown, allowing the clinicians to evaluate it and conclude if there is a luminal obstruction. The presence of obstructive CAD in ICA, usually defined as a lesion greater than 70 percent, has been recognized as an unequivocal sign of a bad cardiovascular prognosis. In contrast, it was initially proposed that non-obstructive CAD (usually defined as a lesion less than 70 percent) could constitute a condition related to a good cardiovascular prognosis \cite{kemp_seven_1986}, but subsequent evidence has increasingly shown that it confers an adverse prognosis when compared to the prognosis in the absence of CAD \cite{rodriguez-capitan_prognostic_2021, wang_prevalence_2017,radico_determinants_2018}. Consequently, it is currently accepted that cardiovascular risk increases when the degree of stenosis increases. However, risk evaluation based on ICA has several limitations. There is enough evidence indicating that the visual assessment of stenosis severity alone has significant interobserver variability, so this visual assessment alone does not provide us with enough information upon which to base decisions about revascularization in many patients \cite{curzen_does_2014}. In addition to this, angiographic assessment of CAD severity is limited in providing consistent information regarding the physiological significance of coronary lesions. Angiography is especially limited in coronary stenoses of intermediate severity (40–70 percent obstruction), where it predicts functional significance in less than 50 percent of lesions \cite{tonino_fractional_2009}. Visual assessment of coronary angiography fails to adequately determine lesion significance because lumen stenosis is only one variable out of many that influence the flow limitation of coronary lesions \cite{topol_our_1995}. Lesion length, collateral flow, and the amount and health of the myocardial bed supplied are other essential factors that are not readily assessed by coronary angiography \cite{halon_can_2018}.  In order to overcome the aforementioned limitations, current guidelines recommend the routine assessment of vessel physiology in the form of indices derived from invasive pressure wire, such as fractional flow reserve and the instantaneous wave-free ratio \cite{knuuti_2019_2020, collet_2020_2021}. Despite these recommendations, the implantation of these functional tests in clinical practice has been especially low \cite{gabara_coronary_2020}.

Many medical image datasets have been provided to the research community with the aim of developing an algorithm that can serve as a computer-aided diagnosis system \cite{tschandl2018ham10000, codella2018skin, spanhol2015dataset, matuszewski2021tem}. However, there is a lack of available and high-quality open-access datasets regarding ICA images because most related studies use private image sets \cite{ovalle2022improving}. Some are provided by an associated medical center and used for image segmentation tasks \cite{nasr2018segmentation, iyer2021angionet}, while others are focused on detection and classification \cite{zhou2021automated, cong2019automated}. None of them provides access to other researchers to their data, which is necessary to achieve advances in this field. The main contributions of this work can be listed as follows:

\begin{itemize}
    \item To provide the research community with a comprehensive and rigorous scientific coronary angiography dataset formed by a set of videos acquired from patients and metadata related to diseases associated with them. This dataset may serve medical doctors to train their skills in angiographic assessment of CAD severity, and computer scientists to create computer-aided diagnosis systems to help with that kind of evaluation and to validate and improve existing methods for CAD detection by training them on more data.
    \item A set of resources for testing algorithms. Additionally, an exhaustive revision and expansion process of these tools will be carried out regularly.
    \item To provide a study of the performance of known architectures using the dataset with the aim of classifying ICA images according to the presence of lesions.
    \item To help the community to identify other related challenges to provide a focus for future research.
\end{itemize}
	
The rest of the paper is structured as follows: Section \ref{sec:related} describes the recent state-of-art works related to ICA. In Section \ref{sec:dataset}, the most important details about the creation and organization of the CADICA dataset are given. The experimental results are shown in Section \ref{sec:results}. A discussion is provided in Section \ref{sec:discussion}. Finally, Section \ref{sec:conclusions} is devoted to conclusions.

\section{Related works}
\label{sec:related}

Deep learning has been thoroughly used for both classification and segmentation tasks in medical imaging, including in the area of cardiology, where the most common imaging modalities are MRI (magnetic resonance imaging), and X-ray-based, such as CCTA (Coronary Computed Tomography Angiography) and ICA (Invasive Coronary Angiography) \cite{song2022deep}.

Specifically for ICA images, we found the work of Au\textit{et al}. \cite{au2018automated}, which uses ICA images of the right coronary artery to detect and classify coronary stenosis. They implemented a complex method based on three phases with different networks to do it.   First, they used a detection network, YOLONet, to localize the patch where is the lesion. Straightaway, the lesion is segmented using the U-Net model, and finally, it is classified by a small CNN as stenosis if the narrowing is higher than 70\%. 

Nasr-Esfahani \textit{et al}. \cite{nasr2018segmentation} used patches from 44 coronary angiographies to set a system, composed of two CNNs based on learned kernels, which returns a segmentation probability map of the ICA images. 

Wu \textit{et al}. \cite{wu2020automatic} proposed a method to localize stenosis lesions in ICA image sequences based on three stages. For this study, 148 sequences of ICA images were employed. Firstly, the most appropriate frames from the complete sequence are selected using the U-Net architecture. Next, a deconvolutional single-shot multibox detector localizes the possible boxes that contain stenosis, and finally, they designed a temporal module to determine which boxes are actually stenosis, considering the temporal sequence of the boxes selected. 

Zhang \textit{et al}. \cite{zhang2022x} used the U-net architecture to implement the C-Unet model that is a deep learning-based solution to automatically extract the centerlines of blood vessels from ICA images, achieving a precision higher than 80\%. 

Cong \textit{et al}. \cite{cong2019automated} implemented a system completely automatically for the classification and location of ICA images, which had been clustered into the left coronary artery and right coronary artery, and the lesions were into categories depending on their grade. To start, an inception-v3 model classified the images according to the projection to which they belonged. Then, the ideal candidate frames were selected by a fusion between inception-v3 and LSTM (long-short-term memory), these candidates were classified into predefined categories by another inception-v3 model adapted for it, and to conclude, they employed a class activation map to identify the regions in ICA images where the lesion could be located. 

Zhou \textit{et al}. \cite{zhou2021automated} proposed a method to classify ICA images with stenosis. To carry out the study, they employed 8731 right coronary artery images, and the procedure was formed by three stages. Firstly, by ResNet-18 structure was carried out to extract keyframes, which presents enough contrast and clarity, from video sequences, there were 6533 non-key frames and 2198 keyframes Secondly, a vessel segmentation was implemented by a U-Net model to get their masks. And thirdly, lesions were measured by the skeletonization of the vessels in masks.

Moon \textit{et al}. \cite{moon2021automatic} classified the ICA images into normal and abnormal arteries if the narrowing is lower or higher than 50\%, respectively. As in the previous work reported, in this study, the first phase was based on extracting key frames from sequences of ICA videos, 542 videos were used, and then, an inceptionv3 architecture was fed with these keyframes to can classify them into normal or abnormal. To finalize, stenosis lesions were visually localized using class activation mapping.

These reported works have some aspects in common with the study that we had been carrying out, but most of them are focused on implementing segmentation or using patches to classify the images, instead of complete images. Also, some of them excluded images with more than one lesion or only classify obstructive lesions, they utilize smaller datasets, most of them are private datasets, the detail of the artery and projections used is omitted or a different annotation is implemented.

\section{CADICA dataset}
\label{sec:dataset}

Next, the most important details about the creation and organization of the CADICA dataset are presented.

\subsection{Patient Selection}
The dataset proposed in this work consists of 668 invasive coronary angiography videos from 42 patients, acquired at Hospital Universitario Virgen de la Victoria, M\'alaga, Spain. They have been included within the regulation set by the local ethical committee of the hospital and patient consent was waived, because this is a retrospective study with anonymized data. Prerequisites and data selection have not been performed in order to impose clinical fidelity. Therefore, a wide variability of cases and acquisition configurations were implied, while some cases were laborious to tag. This way, the dataset exhibits a high variety of pathological cases and image quality. Table \ref{tab:baseline} summarizes the baseline and demographic characteristics of patients included in the dataset.


\begin{table}[ht]
\centering

    \caption{Demographic and baseline characteristics of the patients. Data are given as \% or as median (interquartile range)} 
    \begin{tabular}{|l|c|} 
     \hline
     Age (years) & 71.5 (58.25-78) \\ \hline
        Sex (female-male) & 47.6\% - 52.4\%\\ \hline
        Diabetes mellitus & 40.5\% \\ \hline
        Dyslipidemia & 40.5\% \\ \hline
        Smoker & 45.2\% \\ \hline
        High blood pressure & 61.9\% \\ \hline
        Kidney failure & 14.3\% \\ \hline
        Heart failure & 14.3\% \\ \hline
        Atrial fibrillation & 4.8\% \\ \hline
        \multicolumn{2}{|c|}{Left ventricular ejection fraction}  \\ \hline
        Normal (ejection fraction $>$55\%) & 68.2\% \\ \hline
        Mild dysfunction (ejection fraction 45\%-55\%) & 9.8\% \\ \hline
        Moderate dysfunction (ejection fraction 45\%-35\%) & 0\% \\ \hline
        Severe dysfunction (ejection fraction $<$35\%) & 22\% \\ \hline
        \multicolumn{2}{|c|}{Clinical indication for angiography}  \\ \hline
        Chronic coronary syndrome & 4.9\% \\ \hline
        Non-ST segment elevation acute coronary syndrome & 65.9\% \\ \hline
        ST segment elevation acute coronary syndrome & 29.3\% \\ \hline
        \multicolumn{2}{|c|}{Number of vessels affected}  \\ \hline
        0 & 23.8\% \\ \hline
        1 & 50\% \\ \hline
        2 & 14.3\% \\ \hline
        3 & 11.9\% \\ \hline
        \multicolumn{2}{|c|}{Maximum degree of the coronary artery involvement}  \\ \hline
        $<$20\% & 14.3\% \\ \hline
        20-50\% & 66.7\% \\ \hline
        $>$70\% & 19\% \\ \hline
      
    \end{tabular}
    \label{tab:baseline}
\end{table}

\subsection{Acquisition Protocol}

The invasive coronary angiography videos were acquired as Digital Imaging and Communication in Medicine (DICOM) files recorded at 10 frames per second and with different duration (4-8 seconds) depending on the projection used, but they were converted to PNG images for effortless management. The frame size of each video is 512 $\times$ 512 pixels, while the length of the videos varies from 1 to 151 frames.  The cardiac angiography equipment used was Artis Zee (Siemens AG, Muenchen, Germany). The dose of radiation administered in each projection ranges between 5-50 mGy. The protocol normally used in each angiography included five projections for the left coronary artery (LCA), such as right anterior oblique (RAO) and left anterior oblique (LAO), both with cranial and caudal angulation, with some additional projections in case of diagnostic difficulties. The projections used for the right coronary artery (RCA) are LAO and RAO, with cranial and caudal angulation. 

\begin{figure*}[ht]
	\centering
	\subfigure[LAO 45º CAU 35º]{\includegraphics[width = 2.9 cm]{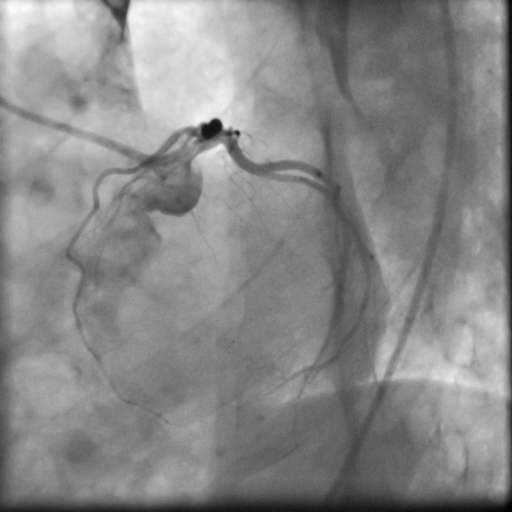}}
	\subfigure[LAO 45º CAU 25º]{\includegraphics[width = 2.9 cm]{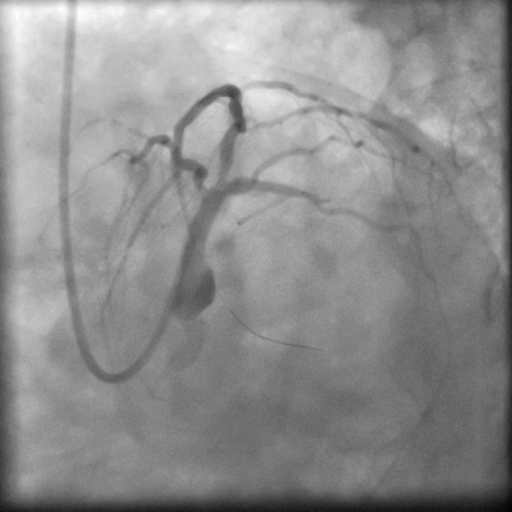}} 
	\subfigure[RAO 30º CRA 20º]{\includegraphics[width = 2.9 cm]{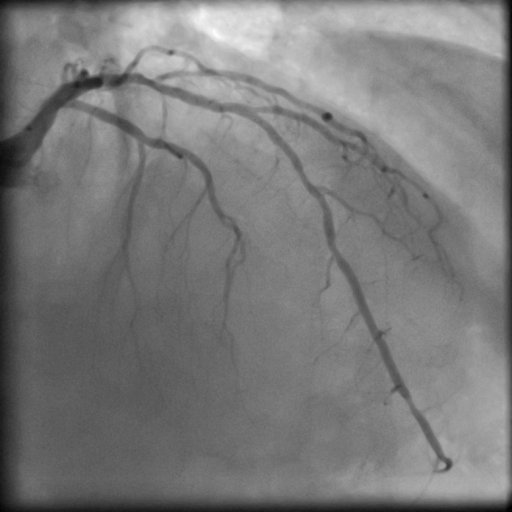}}
	\subfigure[RAO 28º CAU 15º]{\includegraphics[width = 2.9 cm]{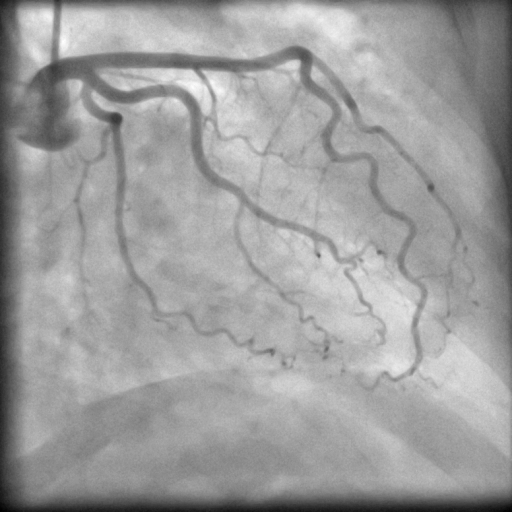}}
	\caption{Examples for left coronary artery (LCA).}
	\label{fig:LCAviews}
\end{figure*}

\begin{figure*}[ht]
	\centering
	\subfigure[LAO 35º CAU 5º]{\includegraphics[width = 2.9 cm]{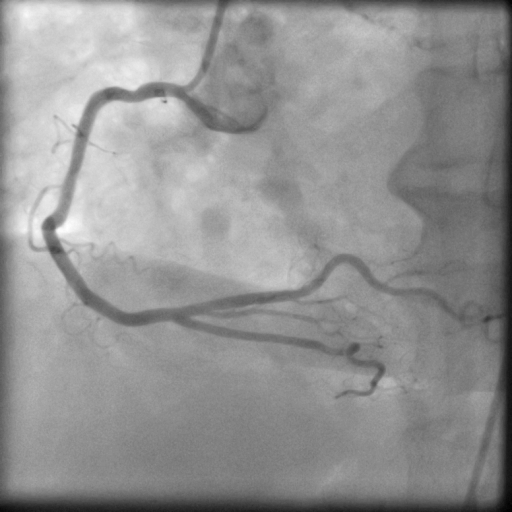}}
	\subfigure[LAO 30º CRA 25º]{\includegraphics[width = 2.9 cm]{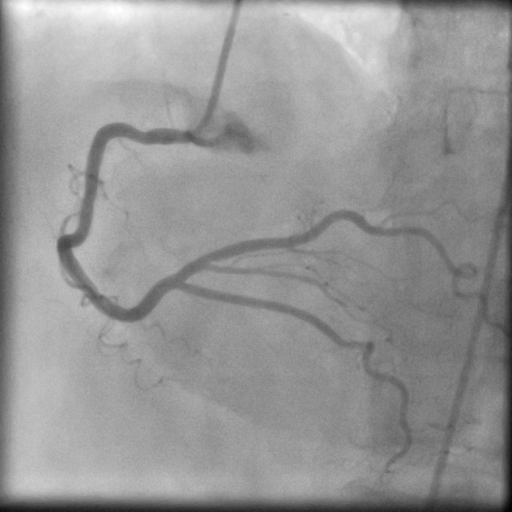}} 
	\subfigure[LAO 25º CRA 20º]{\includegraphics[width = 2.9 cm]{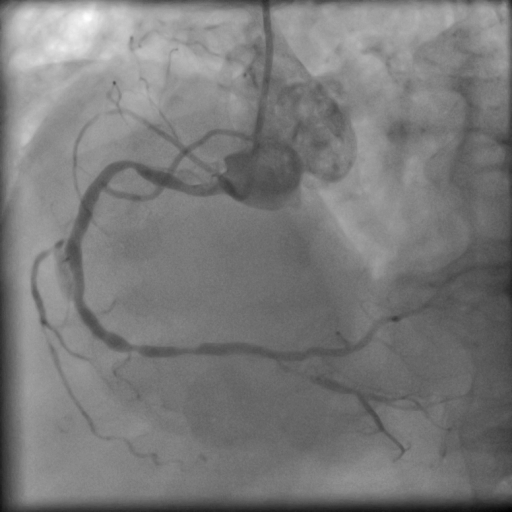}}
	\subfigure[RAO 30º]{\includegraphics[width = 2.9 cm]{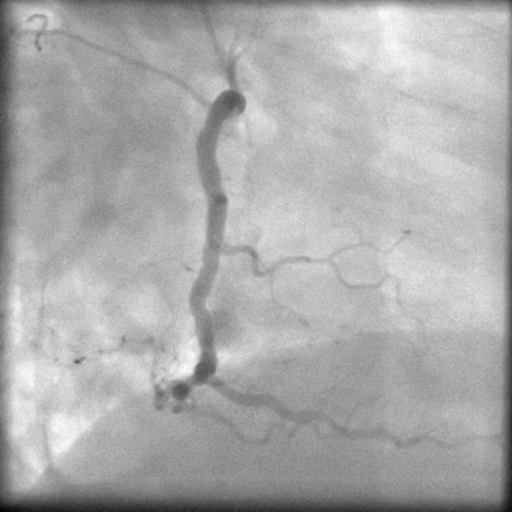}}
	\caption{Examples of projections for right coronary artery (RCA).}
	\label{fig:RCAviews}
\end{figure*}

Fig. \ref{fig:LCAviews} and \ref{fig:RCAviews} show examples of projections for the left coronary artery and the right coronary artery, respectively.

\subsection{Label Protocol}

A team of cardiologists was involved in the annotation of the dataset, assisted by computer scientists. For each frame, those regions of interest are delimited by a bounding box and classified into categories. This way, for each region of interest, is provided the location of the top left corner of its bounding box, its width, its height, and the label of that region of interest. The possible categories are itemized in Table \ref{tab:categories}. 

\begin{table}[ht]
\centering
    \caption{Categories into which lesions have been divided.} 
    \begin{tabular}{|c|c|} 
     \hline
      \textbf{Label} & \textbf{Lesion range} \\ \hline
      p0\_20 & $<$20\%\\ \hline
      p20\_50 & [20\%, 49\%] \\ \hline
      p50\_70 & [50\%, 69\%] \\ \hline 
      p70\_90 & [70\%, 89\%] \\ \hline
      p90\_98 & [90\%, 98\%] \\ \hline
      p99 & 99\% \\ \hline
      p100 & 100\% \\ \hline
      
    \end{tabular}
    \label{tab:categories}
\end{table}

\begin{figure*}[ht]
	\centering
	\subfigure{\includegraphics[width = 3.55 cm]{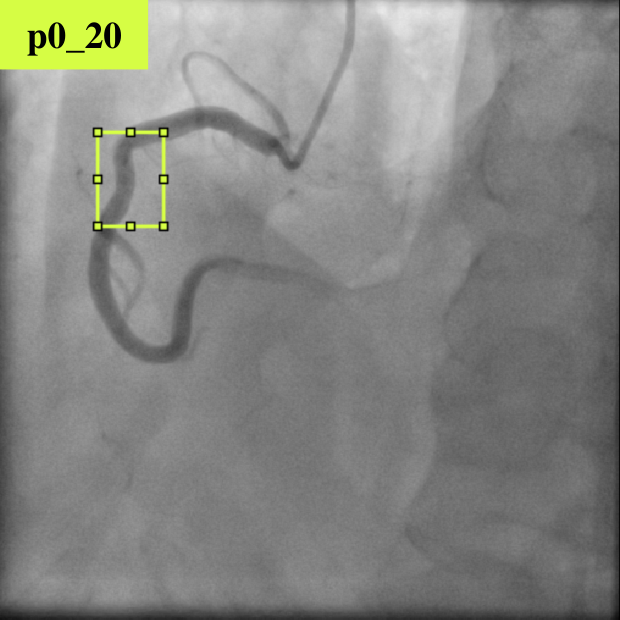}} \qquad
        \subfigure{\includegraphics[width = 3.55 cm]{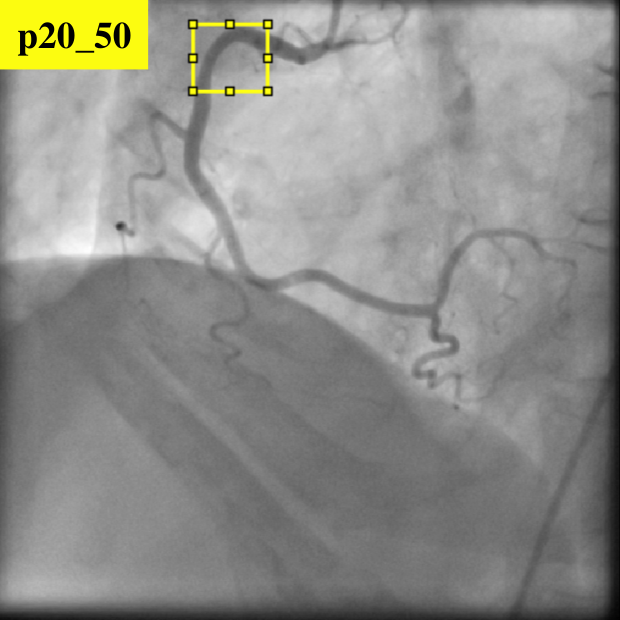}} \qquad
        \subfigure{\includegraphics[width = 3.55 cm]{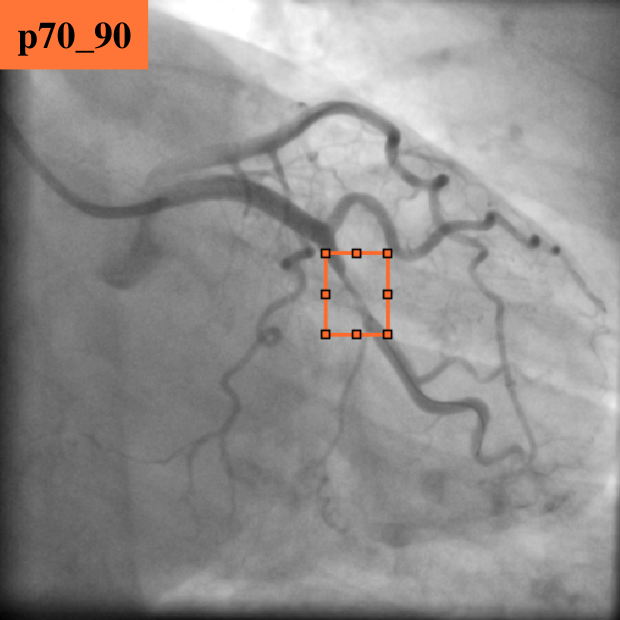}} \qquad
    
        \setcounter{subfigure}{0}
        
	\subfigure[Mild]{\includegraphics[width = 3.55 cm]{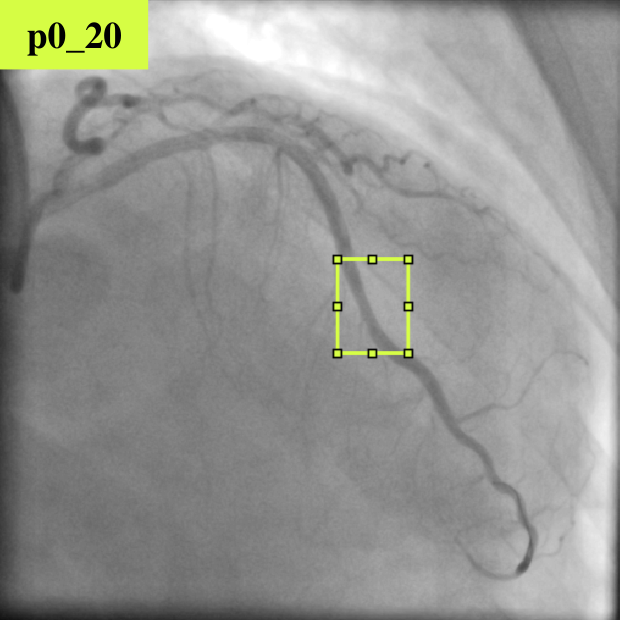}} \qquad
	\subfigure[Moderate]{\includegraphics[width = 3.55 cm]{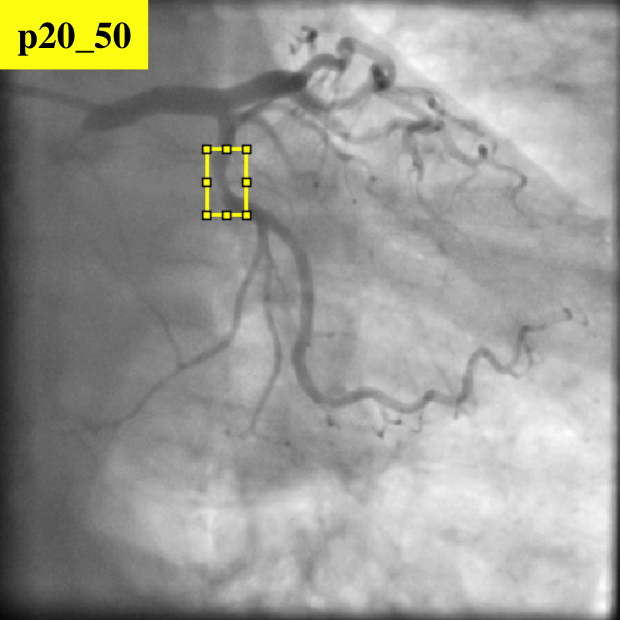}} \qquad
	\subfigure[Severe]{\includegraphics[width = 3.55 cm]{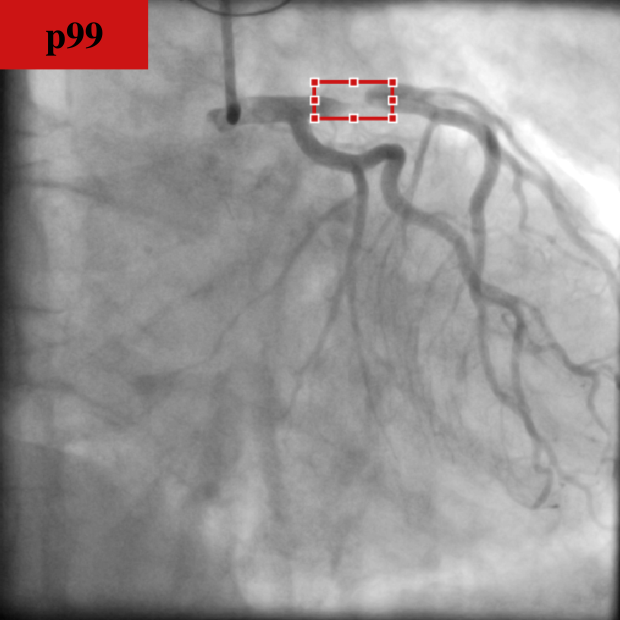}} \qquad
 
	\caption{Samples of the 3 categories in which lesions are classified and delimited by a bounding box annotated.}
	\label{fig:samples}
\end{figure*}

For each video, a selection of keyframes is carried out. This selection contains the list of frames that exhibit a contrast with enough appearance in order to classify the patient correctly. Videos are organized by patients, where a certain number of videos have been collected for each patient. According to their coronary artery stenosis percentage, patients are grouped into three different categories: $<$ 20\% (mild), 20 - 50\% (moderate), and $>$ 70\% (severe). Lesions of 100\% imply a total occlusion of the vessel, while a 99\% lesion presents a gap where the radiocontrast is imperceptible, but the continuation of the vessel is visible. Those lesions that had a narrowing between 50 to 70 percent are classified as obstructive in some studies \cite{finck201910} and non-obstructive in others \cite{kang2016long}, while lesions with a higher narrowing than 70 percent in the previous bibliography showed consensus that they should be taken as obstructive and classify as non-obstructive the lesions that are solidly taken as non-obstructive (20–50 per-cent). Fig. \ref{fig:samples} exhibits some sample images from different patients according to their classification into these categories.

\subsection{Dataset Organization}
\subsubsection{Video Selection}

The presented dataset provides a total of 668 ICA videos. However, in some frames of these videos where CAD is present, the lesion is indiscernible, being difficult to use them for diagnosis. In order to obtain the best videos for CAD classification, a selection of 382 videos was performed. These videos have been chosen by the medical team, where CAD can be visually classified correctly. Thus, videos in which radiocontrast does not perfuse have not been selected for the classification task. The specifications of CADICA dataset are reported in Table \ref{tab:infoCADICA}, where the number of patients, videos, images from selected videos and labels from “lesion” images are itemized. Please note that the number of labels is higher than the number of “lesion” images since it can be more than one lesion in an image.

\begin{table}[ht] 
    \centering
    \caption{Specifications of CADICA dataset: number of patients, videos, images from selected videos and labels from ``lesion'' images.}
    \begin{tabular}{|c|c|c|c|}
    \hline
    \multicolumn{2}{|c|}{\textbf{Patients}} & 42 & \textbf{Total} \\
    \hline
    \multirow{2}{*}{\textbf{Videos}} 
        & Non-Selected & 286 & \multirow{2}{*}{668}\\
        \cline{2-3}
        & Selected & 382 & \\
    \hline
    \multirow{2}{*}{\textbf{Images}} 
        & Non-Lesion & 2,130 & \multirow{2}{*}{6,126} \\ 
        \cline{2-3}
        & Lesion & 3,996 &\\ 
        \hline
    \multirow{7}{*}{\textbf{Labels}} 
        & p0\_20 & 1,944 & \multirow{7}{*}{6,161} \\ 
        \cline{2-3}
        & p20\_50 & 1,128 &\\ 
        \cline{2-3}
        & p50\_70 & 999 &\\ 
        \cline{2-3}
        & p70\_90 & 893 &\\ 
        \cline{2-3}
        & p90\_98 & 930 &\\ 
        \cline{2-3}
        & p99 & 63 &\\ 
        \cline{2-3}
        & p100 & 204 &\\ 
        \hline
    \end{tabular}
    \label{tab:infoCADICA}
\end{table}

\subsubsection{Metadata}
The dataset also provides additional clinical data associated with each patient, such as if the patient suffers from diabetes, dyslipidemia, smoking, hypertension, another comorbidity (such as chronic obstructive pulmonary disease), renal insufficiency, heart failure, atrial fibrillation, or left ventricular ejection fraction. Other information such as age, gender, height, weight, and later event (such as non-cardiac death or heart attack) is also reported.

\subsubsection{Structure}
The CADICA dataset becomes a directory that contains the \textit{metadata.xlsx} file, which is the file where the clinical data is located, as well as two main folders that differentiate the videos selected by the medical team for each patient: \textit{nonselectedVideos} and \textit{selectedVideos}. Inside each folder, there are several sub-directories with the naming convention \textit{pX} where \textit{X} is the ID of each patient, and \textit{vY}, where \textit{Y} is the ID of the video of that patient. 

The folder \textit{pX} contains the following information: 
\begin{itemize}
       \item \textit{vY}: several sub-directories with the videos selected for that patient. 
        \item \textit{lesionVideos.txt}: contains the IDs of the selected videos where appears at least one lesion which is labeled.
       \item \textit{nonlesionVideos.txt}: contains the IDs of the selected videos where there are no visible lesions.
 \end{itemize}

The folder \textit{vY} contains the following information:
 \begin{itemize}
       \item \textit{input}: a sub-directory containing a separate PNG file for each frame of the video.
       \item \textit{pX\_vY\_selectedFrames.txt}: contains the IDs of the keyframes for the medical team, for all the selected videos. 
       \item \textit{groundtruth}: a sub-directory available only if there are lesions in that selected video.
\end{itemize}  

The folder \textit{groundtruth} contains the following information:
 \begin{itemize}
       \item \textit{pX\_vY\_000ZZ.txt}: contains the bounding boxes and their category in each row. There are such files as frames in \textit{pX\_vY\_selectedFrames.txt}. Bounding boxes are specified in the format $[x,y,w,h]$, where $(x,y)$ are the pixel coordinates of the top left corner, $w$ is the width and $h$ is the height of the bounding box.
       \item \textit{pX\_vY\_groundTruthTable.mat}: contains a table with the ground truth information of that video. 
\end{itemize}

\section{Experiments}
\label{sec:results}

Given the provided dataset in this work, we did an exhaustive performance comparison to classify ICA images according to the presence of lesions, being a binary problem, where the method classifies between “non-lesion” or “lesion” (images with any label from Table \ref{tab:categories}).

\subsection{Evaluation metrics}
\label{sec:metrics}

In order to measure the performance of a method that classifies coronary angiography images according to their coronary artery stenosis percentage, several well-known metrics have been proposed.

Let us consider the true positives or number of hits (TP), true negatives or correct rejections (TN), false negatives or misses (FN), and false positives or false alarms (FP). The selected metrics and their definitions are as follows:
\begin{equation}
Acc = \frac{TP + TN}{TP + FP + FN + TN} 
\quad
Fm = 2 \cdot \frac{PR \cdot RC}{PR+RC} 
\end{equation}
\begin{equation}
Bal = \frac{RC + SP}{2}
\qquad
SP = \frac{TN}{FP + TN}
\end{equation}
\begin{equation}
RC = \frac{TP}{TP + FN}
\qquad
PR = \frac{TP}{TP + FP}
\end{equation}

The most representative measures are the Accuracy (Acc), the F-measure (Fm, also known as F1 score), and the Balanced Accuracy (Bal), which provide a good overall evaluation of the performance of a given method. All these measures represent the percentage of hits of the system by providing values in the interval $[0,1]$, where higher is better. 

Meanwhile, other measures are also implicitly considered such as the precision (PR), the recall (RC), and the specificity (SP). In order to analyze these metrics, FN must be considered against FP (lower is better), PR against RC (higher is better). 

\subsection{Methods}
\subsubsection{Convolutional Neural Networks}
In this study, different Convolutional Neural Networks (CNNs) are used to compare their performance to classify ICA images from CADICA into two classes, ``lesion'', which means that appears at least one lesion, and ``non-lesion''. 

CNN is a type of deep learning model incorporating at least one convolutional layer, whose purpose is to extract the features from the input image, triggering under a specific condition \cite{Wang2021}. CNNs have become successful methods with great versatility of applications in several areas, including medical images, and to solve different problems, such as segmentation, localization, or classification. Also, CNNs are characterized by their transferability of knowledge by applying the transfer learning technique, which is based on employing classification models trained on large datasets, also named pre-trained networks, which are re-trained with a specific dataset to specialize them to the particular problem \cite{ovalle2020}. In this study no layer was frozen, so all weights were updated according to the input dataset information. Five known pre-trained CNN architectures are used in this study:

The Residual Networks (ResNets) family \cite{he2016} introduces the residual connection to the model, these shortcut connections allow skipping some layers in the process. In particular, in this study ResNet-18 and ResNet-50 networks are used, which are characterized by being composed of 18 and 50 layers deep, respectively.

MobileNet-V2 \cite{sandler2018} is a mobile neural network optimized to considerably reduce the number of parameters, compared with other architectures, which decreases the computational load. The MobileNet architecture is based on depthwise separable convolutions, which are a combination of two layers. The first is depthwise convolution, which applies a single filter to the input without extracting features, and the second is named pointwise convolution, which creates a linear combination output with new features \cite{sae2019}.

NasNet-Mobile is the smallest version of NasNet models. NasNet models are CNNs based on Neural Architecture Search (NAS), which consist of basic building blocks, called cells, optimized by reinforcement learning method \cite{reddy2018}.

DenseNet-201 is a deep network based on ensuring the maximum information flow between layers by dense blocks. Dense blocks are blocks of layers, where each layer is connected to all former layers, instead only to the previous one \cite{huang2017}.

\subsubsection{Data Preprocessing}

In CADICA there are 382 selected videos in total, of which two subgroups were done differentiating between views of the left (LCA) and right (RCA) coronary arteries.

The subgroup of LCA views was composed of 216 videos, a total of 3,228 images, where 1,003 images were labeled as ``non-lesion'' and 2,225 were labeled as ``lesion''. The subgroup of RCA views consisted of 118 videos, which is 2,077 images in total, whose labels were distributed as 617 images labeled as ``non-lesion'' and 1,460 labeled as ``lesion''.

The input image of the pre-trained architectures selected is an RGB image of size 224 $\times$ 224 pixels. Thus, the first processing applied to all images was to resize them and use the color preprocessing to ensure that images have the number of channels required, in this case, three channels.

To study the binary classification problem ``lesion''/``non-lesion'', both sets had been divided into training (80\%) and test (20\%) sets. This division was done by videos, which means that 80\% of the ``non-lesion'' and ``lesion'' videos were used for training and the 20\% remaining for testing. This way, frames of the same video of the train set are unavailable for the test set, because frames of a video are very similar between them.

\begin{figure*}[ht]
	\centering
	\subfigure[Original]{\includegraphics[width = 2.9 cm]{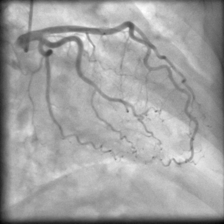}}
    \subfigure[Translation]{\includegraphics[width = 2.9 cm]{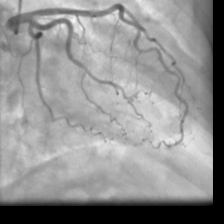}\label{fig:samplesAug_tras}}     
	\subfigure[Rotation]{\includegraphics[width = 2.9 cm]{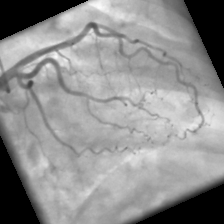}\label{fig:samplesAug_rot}}
	\subfigure[Scale]{\includegraphics[width = 2.9 cm]{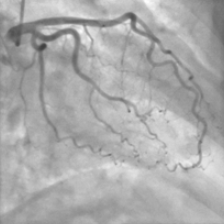}\label{fig:samplesAug_sca}}
	\caption{Examples of the modifications applied to the training sets to augment data.}
	\label{fig:samplesAug}
\end{figure*}

Both sets have unbalanced distributions, which can cause the model to specialize in the majority class, in this case, ``lesion'', and be relatively inefficient at classifying the minority class, in this case, ``non-lesion''.  To solve this issue a data augmentation strategy had been applied to the training sets. This data augmentation was done by using different random basic operations of the original images, detailed as follows:

\begin{itemize}
    \item Translations in the x and y axis of [-25,25] pixels randomly, Fig. \ref{fig:samplesAug_tras}.
    \item Rotation using a random angle between [-25º,25º], Fig. \ref{fig:samplesAug_rot}.
    \item Scaling of the images with a random scale factor in a range of [0.8,1.7], Fig.  \ref{fig:samplesAug_sca}. 
\end{itemize}

\begin{figure}[ht]
	\centering
	\includegraphics[scale=0.4]{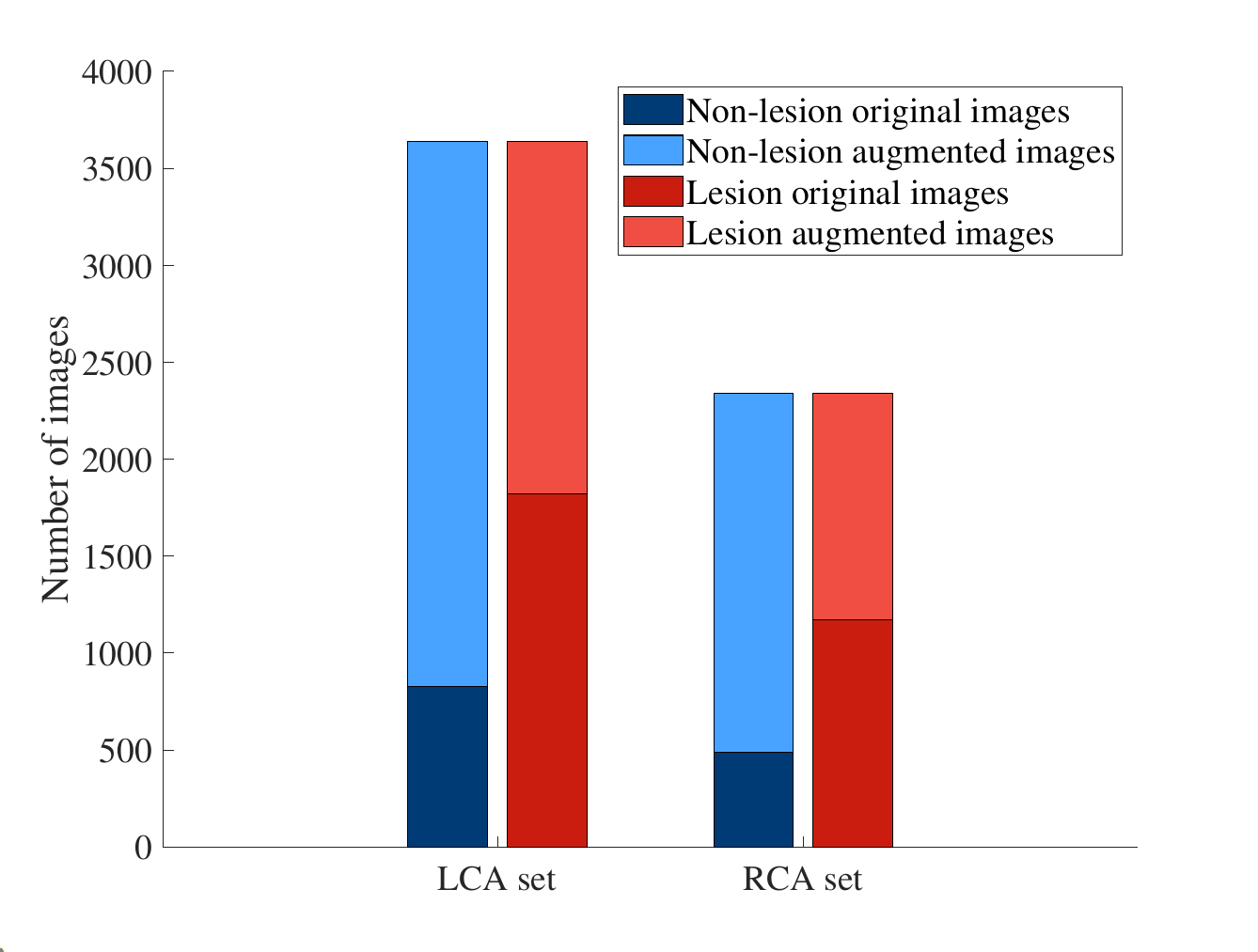}
	\caption{Class distribution for Left coronary Artery (LCA) and Right Coronary Artery (RCA) sets.}
	\label{fig:classDistribution}
\end{figure}

Modifications were applied to the training sets of both classes, ``lesion'' and ``non-lesion'' images, and to both subsets, LCA and RCA, equalizing and increasing them. Fig.  \ref{fig:classDistribution} shows the original class distributions for LCA and RCA sets, and the final distributions obtained after data augmentation was implemented, obtaining 3640 and 2342 images of each class in the LCA set and in the RCA set, respectively.

\subsubsection{Experimental Setup}
Several parameters can be tuned for training convolutional neural networks. The main ones that we were focusing on are reported below:

\begin{itemize}
    \item Validation frequency: it is the number of iterations between evaluations of the training process.
    \item Maximum number of epochs: indicates the maximum times that the full dataset is passed to the model to update its weights.
    \item Optimizer or solver: is the algorithm applied to update the weights of the network to reduce the loss function.
    \item Initial Learning Rate: establishes the rate that is going to use to start the learning procedure.
    \item Batch size: specifies the number of samples, in this case, images, that are processed by the model in one iteration.
\end{itemize}
	
Due to many options for possible combinations of training parameters, we started studying the behavior of the performance, establishing some values to tune the training parameters. To evaluate the progress in the tuning of the parameters, we used the performance metrics reported in section \ref{sec:metrics}, focusing on F-measure, Balanced Accuracy, and Accuracy, because together they provide a global view of the performance. The LCA set was employed for this process because it is the largest and most complex set. Finally, the parameters selected were used to evaluate the RCA set too.

The first parameters that we set were validation frequency in 50 iterations to evaluate the training process and the maximum number of epochs established in 10 epochs because the increase of it had an unsubstantial improvement compared to time-consuming. For the optimizer, we compared different algorithms: Adam (adaptive moment estimation), SGDM (stochastic gradient descent with momentum), and RMSProp (root mean square propagation). Besides, we proposed several rates for the initial learning rate: 0.01, 0.001, 0.0001, and 0.00001. For the batch size, two values were selected, 16 and 64. In total there are 24 possible combinations, to compare them, we implemented the stratified K-fold cross-validation. This technique is adequate to evaluate the performance in a reliable way because the results are averages from different partitions of the input dataset, so results are independent of the partition employed to validate.

\begin{table*}[ht] 
    \centering
    \caption{F-measure results obtained on the test set for LCA images using 5-fold stratified cross-validation, five convolutional network architectures, and different values for batch size, initial learning rate, and optimizer. The best performances are shown in bold.} 
    \resizebox{\textwidth}{!}{
    \begin{tabular}{|l|l|ccc|ccc|}
    \hline
     \multirow{2}{*}{Model} & \multirow{2}{*}{\shortstack[l]{Initial\\ Learning Rate}} & \multicolumn{3}{c|}{Batch Size 16} & \multicolumn{3}{c|}{Batch Size 64}  \\ 
    \cline{3-8} 
        &  & \emph{Adam} & \emph{SGDM} & \emph{RMSProp} & \emph{Adam} & \emph{SGDM} & \emph{RMSProp} \\ 
        \hline
        \multirow{4}{*}{\emph{MobileNet-V2}} 
         & 0.01 & $0.758 \pm 0.033$ & $0.776 \pm 0.028$ & $0.755 \pm 0.096$ & $0.767 \pm 0.036$ & $0.806 \pm 0.013$ & $0.720 \pm 0.052$   \\ 
         & 0.001 & $0.798 \pm 0.035$ & $0.800 \pm 0.025$ & $0.791 \pm 0.034$ & $0.812 \pm 0.024$ & $0.786 \pm 0.018$ & $0.807 \pm 0.024$   \\ 
         & 0.0001 & $0.809 \pm 0.030$ & $0.794 \pm 0.021$ & $0.818 \pm 0.015$ & $0.806 \pm 0.016$ & $0.800 \pm 0.010$ & $0.808 \pm 0.015$   \\ 
         & 0.00001 & $0.810 \pm 0.014$ & $0.785 \pm 0.010$ & $\boldsymbol{0.820 \pm 0.015}$ & $\boldsymbol{0.810 \pm 0.009}$ & $0.766 \pm 0.011$ & $0.789 \pm 0.012$   \\ 
         \hline

        \multirow{4}{*}{\emph{ResNet-18}} 
         & 0.01 & $0.737 \pm 0.054$ & $0.739 \pm 0.038$ & $0.730 \pm 0.094$ & $0.747 \pm 0.047$ & $0.789 \pm 0.011$ & $0.731 \pm 0.065$   \\ 
         & 0.001 & $0.782 \pm 0.020$ & $0.798 \pm 0.020$ & $0.765 \pm 0.025$ & $0.784 \pm 0.013$ & $\boldsymbol{0.794 \pm 0.016}$ & $0.762 \pm 0.017$   \\ 
         & 0.0001 & $0.784 \pm 0.022$ & $0.780 \pm 0.022$ & $\boldsymbol{0.800 \pm 0.021}$ & $0.783 \pm 0.029$ & $0.788 \pm 0.026$ & $0.775 \pm 0.014$   \\ 
         & 0.00001 & $0.783 \pm 0.017$ & $0.800 \pm 0.029$ & $0.789 \pm 0.019$ & $0.772 \pm 0.016$ & $0.770 \pm 0.016$ & $0.782 \pm 0.012$   \\ 
         \hline

        \multirow{4}{*}{\emph{ResNet-50}}
         & 0.01 & $0.655 \pm 0.057$ & $0.777 \pm 0.027$ & $0.380 \pm 0.418$ & $0.755 \pm 0.006$ & $0.793 \pm 0.013$ & $0.713 \pm 0.097$   \\ 
         & 0.001 & $0.773 \pm 0.020$ & $0.774 \pm 0.012$ & $0.775 \pm 0.031$ & $0.772 \pm 0.042$ & $0.779 \pm 0.021$ & $0.746 \pm 0.055$   \\ 
         & 0.0001 & $0.785 \pm 0.023$ & $0.784 \pm 0.025$ & $\boldsymbol{0.823 \pm 0.017}$ & $0.806 \pm 0.024$ & $0.802 \pm 0.016$ & $\boldsymbol{0.818 \pm 0.023}$   \\ 
         & 0.00001 & $0.797 \pm 0.037$ & $0.796 \pm 0.026$ & $0.808 \pm 0.025$ & $0.782 \pm 0.040$ & $0.783 \pm 0.034$ & $0.777 \pm 0.024$   \\ 
         \hline

        \multirow{4}{*}{\emph{NasNet-Mobile}}
         & 0.01 & $0.743 \pm 0.018$ & $0.785 \pm 0.017$ & $0.712 \pm 0.105$ & $0.723 \pm 0.037$ & $\boldsymbol{0.798 \pm 0.011}$ & $0.717 \pm 0.047$   \\ 
         & 0.001 & $0.807 \pm 0.026$ & $\boldsymbol{0.811 \pm 0.028}$ & $0.790 \pm 0.032$ & $0.790 \pm 0.027$ & $0.785 \pm 0.019$ & $0.753 \pm 0.050$   \\ 
         & 0.0001 & $0.794 \pm 0.025$ & $0.770 \pm 0.036$ & $0.803 \pm 0.017$ & $0.779 \pm 0.015$ & $0.750 \pm 0.021$ & $0.783 \pm 0.016$   \\ 
         & 0.00001 & $0.778 \pm 0.008$ & $0.714 \pm 0.011$ & $0.770 \pm 0.016$ & $0.773 \pm 0.006$ & $0.659 \pm 0.049$ & $0.765 \pm 0.010$   \\ 
         \hline
        
        \multirow{4}{*}{\emph{DenseNet-201}}
         & 0.01 & $0.694 \pm 0.020$ & $0.736 \pm 0.032$ & $0.666 \pm 0.103$ &  &  &    \\ 
         & 0.001 & $0.761 \pm 0.032$ & $0.799 \pm 0.019$ & $0.758 \pm 0.025$ &  &  &     \\ 
         & 0.0001 & $0.797 \pm 0.036$ & $0.774 \pm 0.012$ & $0.801 \pm 0.034$ &  &  &     \\ 
         & 0.00001 & $0.806 \pm 0.025$ & $\boldsymbol{0.826 \pm 0.017}$ & $0.794 \pm 0.014$ &  &  &    \\ 
        \hline
    
    \end{tabular}
    }\label{tab:5fold-fscore}
\end{table*}


 \begin{table*}[ht] 
     \centering
    \caption{Balanced Accuracy results obtained on the test set for LCA images using 5-fold stratified cross-validation, five convolutional network architectures, and different values for batch size, initial learning rate, and optimizer. The best performances are shown in bold.} 
    \resizebox{\textwidth}{!}{
    \begin{tabular}{|l|l|ccc|ccc|}
     \hline
     \multirow{2}{*}{Model} & \multirow{2}{*}{\shortstack[l]{Initial\\ Learning Rate}} & \multicolumn{3}{c|}{Batch Size 16} & \multicolumn{3}{c|}{Batch Size 64}  \\
     \cline{3-8}
        &  & \emph{Adam} & \emph{SGDM} & \emph{RMSProp} & \emph{Adam} & \emph{SGDM} & \emph{RMSProp} \\ 
        \hline
        \multirow{4}{*}{\emph{MobileNet-V2}} 
         & 0.01 & $0.586 \pm 0.059$ & $0.597 \pm 0.052$ & $0.514 \pm 0.029$ & $0.608 \pm 0.029$ & $0.610 \pm 0.024$ & $0.578 \pm 0.041$ \\ 
         & 0.001 & $0.656 \pm 0.040$ & $0.644 \pm 0.052$ & $0.650 \pm 0.044$ & $0.660 \pm 0.073$ & $0.630 \pm 0.022$ & $0.648 \pm 0.052$ \\ 
         & 0.0001 & $0.667 \pm 0.034$ & $0.643 \pm 0.035$ & $0.671 \pm 0.011$ & $0.665 \pm 0.014$ & $0.638 \pm 0.030$ & $0.657 \pm 0.018$ \\ 
         & 0.00001 & $0.682 \pm 0.020$ & $0.618 \pm 0.013$ & $\boldsymbol{0.688 \pm 0.024}$ & $\boldsymbol{0.678 \pm 0.027}$ & $0.590 \pm 0.020$ & $0.662 \pm 0.012$ \\ 
         \hline

        \multirow{4}{*}{\emph{ResNet-18}} 
         & 0.01 & $0.551 \pm 0.071$ & $0.559 \pm 0.028$ & $0.492 \pm 0.034$ & $0.594 \pm 0.053$ & $0.625 \pm 0.028$ & $0.582 \pm 0.120$ \\ 
         & 0.001 & $0.609 \pm 0.037$ & $0.618 \pm 0.023$ & $0.605 \pm 0.017$ & $0.636 \pm 0.050$ & $\boldsymbol{0.656 \pm 0.040}$ & $0.583 \pm 0.033$ \\ 
         & 0.0001 & $0.592 \pm 0.014$ & $0.633 \pm 0.027$ & $0.623 \pm 0.017$ & $0.610 \pm 0.037$ & $0.632 \pm 0.037$ & $0.583 \pm 0.027$ \\ 
         & 0.00001 & $0.640 \pm 0.009$ & $0.633 \pm 0.061$ & $\boldsymbol{0.649 \pm 0.024}$ & $0.615 \pm 0.025$ & $0.618 \pm 0.043$ & $0.638 \pm 0.014$ \\ 
         \hline

        \multirow{4}{*}{\emph{ResNet-50}}
         & 0.01 & $0.558 \pm 0.026$ & $0.592 \pm 0.029$ & $0.500 \pm 0.000$ & $0.578 \pm 0.050$ & $0.632 \pm 0.027$ & $0.547 \pm 0.084$ \\ 
         & 0.001 & $0.600 \pm 0.022$ & $0.613 \pm 0.022$ & $0.564 \pm 0.055$ & $0.626 \pm 0.014$ & $0.642 \pm 0.034$ & $0.577 \pm 0.037$ \\ 
         & 0.0001 & $0.636 \pm 0.031$ & $0.653 \pm 0.033$ & $0.645 \pm 0.041$ & $0.638 \pm 0.057$ & $\boldsymbol{0.651 \pm 0.022}$ & $0.648 \pm 0.030$ \\ 
         & 0.00001 & $0.659 \pm 0.060$ & $0.623 \pm 0.039$ & $\boldsymbol{0.679 \pm 0.033}$ & $0.650 \pm 0.047$ & $0.600 \pm 0.066$ & $0.642 \pm 0.023$ \\ 
         \hline

        \multirow{4}{*}{\emph{NasNet-Mobile}}
         & 0.01 & $0.562 \pm 0.020$ & $0.620 \pm 0.048$ & $0.531 \pm 0.039$ & $0.568 \pm 0.043$ & $0.639 \pm 0.018$ & $0.597 \pm 0.024$ \\ 
         & 0.001 & $0.626 \pm 0.045$ & $0.686 \pm 0.034$ & $0.649 \pm 0.053$ & $0.610 \pm 0.047$ & $\boldsymbol{0.647 \pm 0.034}$ & $0.597 \pm 0.049$ \\ 
         & 0.0001 & $0.625 \pm 0.029$ & $0.644 \pm 0.048$ & $\boldsymbol{0.649 \pm 0.025}$ & $0.637 \pm 0.033$ & $0.618 \pm 0.035$ & $0.629\pm 0.017$ \\ 
         & 0.00001 & $0.621 \pm 0.010$ & $0.583 \pm 0.025$ & $0.629 \pm 0.010$ & $0.620 \pm 0.013$ & $0.529 \pm 0.022$ & $0.607 \pm 0.017$ \\ 
         \hline
        
        \multirow{4}{*}{\emph{DenseNet-201}}
         & 0.01 & $0.537 \pm 0.084$ & $0.597 \pm 0.067$ & $0.512 \pm 0.056$ &  &  &  \\ 
         & 0.001 & $0.576 \pm 0.041$ & $0.666 \pm 0.020$ & $0.611 \pm 0.033$ &  &  &   \\ 
         & 0.0001 & $0.623 \pm 0.060$ & $0.627 \pm 0.022$ & $0.669 \pm 0.059$ &  &  &   \\ 
         & 0.00001 & $0.642 \pm 0.037$ & $\boldsymbol{0.672 \pm 0.025}$ & $0.638 \pm 0.022$ &  &  &   \\ 
        \hline
        
    \end{tabular}
    }\label{tab:5fold-balanced}
\end{table*}

\begin{table*}[ht]
     \centering
    \caption{Accuracy results obtained on the test set for LCA images using 5-fold stratified cross-validation, five convolutional network architectures, and different values for batch size, initial learning rate, and optimizer. The best performances are shown in bold.} 
    \resizebox{\textwidth}{!}{
    \begin{tabular}{|l|l|ccc|ccc|}
     \hline
     \multirow{2}{*}{Model} & \multirow{2}{*}{\shortstack[l]{Initial\\ Learning Rate}} & \multicolumn{3}{c|}{Batch Size 16} & \multicolumn{3}{c|}{Batch Size 64}  \\
     \cline{3-8}
        &  & \emph{Adam} & \emph{SGDM} & \emph{RMSProp} & \emph{Adam} & \emph{SGDM} & \emph{RMSProp} \\ \hline 
        \multirow{4}{*}{\emph{MobileNet-V2}} 
         & 0.01 & $0.659 \pm 0.037$ & $0.678 \pm 0.041$ & $0.642 \pm 0.074$ & $0.674 \pm 0.035$ & $0.710 \pm 0.017$ & $0.626 \pm 0.045$ \\ 
         & 0.001 & $0.716 \pm 0.040$ & $0.713 \pm 0.039$ & $0.708 \pm 0.044$ & $0.730 \pm 0.041$ & $0.696 \pm 0.018$ & $0.722 \pm 0.033$ \\ 
         & 0.0001 & $0.729 \pm 0.036$ & $0.708 \pm 0.029$ & $0.739 \pm 0.017$ & $0.726 \pm 0.019$ & $0.712 \pm 0.018$ & $0.725 \pm 0.017$ \\ 
         & 0.00001 & $0.733 \pm 0.018$ & $0.692 \pm 0.012$ & $\boldsymbol{0.745 \pm 0.021}$ & $\boldsymbol{0.732 \pm 0.013}$ & $0.667 \pm 0.014$ & $0.708 \pm 0.012$ \\ \hline 

        \multirow{4}{*}{\emph{ResNet-18}} 
         & 0.01 & $0.629 \pm 0.069$ & $0.634 \pm 0.037$ & $0.611 \pm 0.089$ & $0.653 \pm 0.055$ & $0.698 \pm 0.014$ & $0.634 \pm 0.091$ \\ 
         & 0.001 & $0.687 \pm 0.021$ & $0.705 \pm 0.025$ & $0.671 \pm 0.022$ & $0.697 \pm 0.023$ & $\boldsymbol{0.712 \pm 0.025}$ & $0.661 \pm 0.017$ \\ 
         & 0.0001 & $0.684 \pm 0.021$ & $0.692 \pm 0.027$ & $0.708 \pm 0.023$ & $0.688 \pm 0.036$ & $0.699 \pm 0.033$ & $0.673 \pm 0.019$ \\ 
         & 0.00001 & $0.697 \pm 0.018$ & $\boldsymbol{0.710 \pm 0.043}$ & $0.705 \pm 0.022$ & $0.679 \pm 0.021$ & $0.679 \pm 0.026$ & $0.695 \pm 0.015$ \\ \hline 

        \multirow{4}{*}{\emph{ResNet-50}}
         & 0.01 & $0.572 \pm 0.030$ & $0.678 \pm 0.030$ & $0.473 \pm 0.209$ & $0.653 \pm 0.017$ & $0.703 \pm 0.019$ & $0.617 \pm 0.064$ \\ 
         & 0.001 & $0.676 \pm 0.023$ & $0.681 \pm 0.017$ & $0.650 \pm 0.027$ & $0.685 \pm 0.043$  & $0.694 \pm 0.029$ & $0.646 \pm 0.570$ \\ 
         & 0.0001 & $0.698 \pm 0.022$ & $0.702 \pm 0.030$ & $\boldsymbol{0.737 \pm 0.027}$ & $0.717 \pm 0.038$ & $0.718 \pm 0.019$ & $\boldsymbol{0.732 \pm 0.029}$ \\ 
         & 0.00001 & $0.715 \pm 0.050$ & $0.705 \pm 0.031$ & $0.731 \pm 0.032$ & $0.699 \pm 0.049$ & $0.685 \pm 0.050$ & $0.692 \pm 0.028$ \\ \hline 

        \multirow{4}{*}{\emph{NasNet-Mobile}}
         & 0.01 & $0.638 \pm 0.016$ & $0.692 \pm 0.028$ & $0.613 \pm 0.074$ & $0.623 \pm 0.042$ & $\boldsymbol{0.711 \pm 0.015}$ & $0.630 \pm 0.037$ \\ 
         & 0.001 & $0.715 \pm 0.037$ & $\boldsymbol{0.736 \pm 0.035}$ & $0.706 \pm 0.040$ & $0.695 \pm 0.039$ & $0.701 \pm 0.025$ & $0.659 \pm 0.056$ \\ 
         & 0.0001 & $0.704 \pm 0.028$ & $0.687 \pm 0.044$ & $0.718 \pm 0.021$ & $0.693 \pm 0.021$ & $0.661 \pm 0.028$ & $0.694 \pm 0.018$ \\ 
         & 0.00001 & $0.687 \pm 0.008$ & $0.621 \pm 0.017$ & $0.682 \pm 0.016$ & $0.682 \pm 0.007$ & $0.562 \pm 0.038$ & $0.670 \pm 0.014$ \\ \hline 
        
        \multirow{4}{*}{\emph{DenseNet-201}}
         & 0.01 & $0.589 \pm 0.041$ & $0.643 \pm 0.047$ & $0.569 \pm 0.081$ &  &  &  \\ 
         & 0.001 & $0.658 \pm 0.039$ & $0.719 \pm 0.020$ & $0.666 \pm 0.030$ &  &  &   \\ 
         & 0.0001 & $0.705 \pm 0.049$ & $0.685 \pm 0.015$ & $0.723 \pm 0.032$ &  &  &   \\ 
         & 0.00001 & $0.719 \pm 0.034$ & $\boldsymbol{0.746 \pm 0.024}$ & $0.706 \pm 0.019$ &  &  &   \\  
         \hline
   
    \end{tabular}
    } \label{tab:5fold-accuracy}
\end{table*}

The proposed models were implemented in MATLAB R2022b on a computer system with an Intel Core i9-10900X processor, 128 GB of RAM, and NVIDIA GeForce RTX 3080 Ti GPU card.

Firstly, we divided the training set of LCA subset into 5 boxes with samples of both classes, 80\% for training the models and 20\% as the validation set. For each model, the 24 combinations established were executed, except DenseNet-201, which was only trained with batch size 16 because of memory settings.

The obtained results for test sets, with the different possible combinations for each architecture, are shown in Tables \ref{tab:5fold-fscore}, \ref{tab:5fold-balanced}, and \ref{tab:5fold-accuracy}, where F-measure, Balanced Accuracy and Accuracy values for each combination are reported. In these tables, the best results for each model and batch size are shown in bold. 

By observing these tables, we can see how in general terms is better to choose low learning rates and a batch size of 16 instead of 64. Moreover, SGDM and RMSProp solvers obtained better results than Adam, especially for batch sizes of 16, where the results with the Adam solver never overcome SGDM and RMSProp solvers.

The configurations selected for each model are reported in Table \ref{tab:configurations}. These configurations were chosen according to the results obtained in Tables \ref{tab:5fold-fscore}, \ref{tab:5fold-balanced}, and \ref{tab:5fold-accuracy} with 5-fold cross-validation. The optimizer and the initial learning rate that more times get the highest results were selected. For instance, the best values for MobileNet-V2 with a batch size of 16 were achieved using the RMSProp solver and an initial learning rate of 0.00001 in the three tables. However, some models present different best configurations depending on the measure taken into account, e.g., ResNet-18 with a batch size of 16 obtained the best results using the RMSProp optimizer for F-measure and Balanced Accuracy, whereas the best optimizer was SGDM for Accuracy. In this case, the RMSProp optimizer is selected. Likewise, the chosen initial learning rate was 0.00001 for ResNet-18.


\begin{table}[ht] 
    \centering
    \caption{Selected configurations considering F-measure, Balanced Accuracy and Accuracy obtained with 5-fold stratified cross-validation.}
    \begin{tabular}{|l|c|l|l|} 
    \hline
     Model & Batch Size & Optimizer & Initial Learning Rate \\
    \cline{1-4} 
        \multirow{2}{*}{\emph{MobileNet-V2}} 
         & 16 & RMSprop & 0.00001  \\ 
         & 64 & Adam & 0.00001 \\ 
         \hline 

         \multirow{2}{*}{\emph{ResNet-18}} 
         & 16 & RMSprop & 0.00001 \\ 
         & 64 & SGDM & 0.001 \\ 
         \hline 

         \multirow{2}{*}{\emph{ResNet-50}} 
         & 16 & RMSprop & 0.00001 \\ 
         & 64 & RMSprop & 0.0001 \\ 
         \hline 

         \multirow{2}{*}{\emph{NasNet-Mobile}} 
         & 16 & SGDM & 0.001 \\ 
         & 64 & SGDM & 0.01 \\ 
         \hline 

         \emph{DenseNet-201}
         & 16 & SGDM & 0.00001 \\ 
        \hline
    
    \end{tabular}
    \label{tab:configurations}
\end{table}

\subsection{Results}
The experiments carried out to compare the performance of the different pre-trained models to evaluate the functionality of the implemented dataset based on the selected configurations (Table \ref{tab:configurations}) are summarized here. These configurations were employed to implement a 10-fold stratified cross-validation, where 90\% of images were used for training the models and 10\% to validate the training process. The experimental results using different neural models, batch size and image subgroups (LCA and RCA) are shown in Table \ref{tab:10fold}, where Balanced accuracy, F-measure and Accuracy obtained in the test set are reported.  In Table \ref{tab:10fold} are shown in bold the highest values obtained for each measure in both subsets. Note that the best results for Balanced Accuracy and Accuracy were obtained with the MobileNet-V2 model for LCA subset (0.673 and 0.732, respectively). However, for the RCA set the NasNet-Mobile model achieved the best Balanced Accuracy and Accuracy (0.658, and 0.744). According to the F-measure, the best results were obtained by the ResNet-50 in both subsets (0.814, and 0.830, respectively).

\begin{figure*}[ht]
	\centering
	\subfigure[Ranking of LCA set with batch size 16.]
 {\includegraphics[scale=0.255]{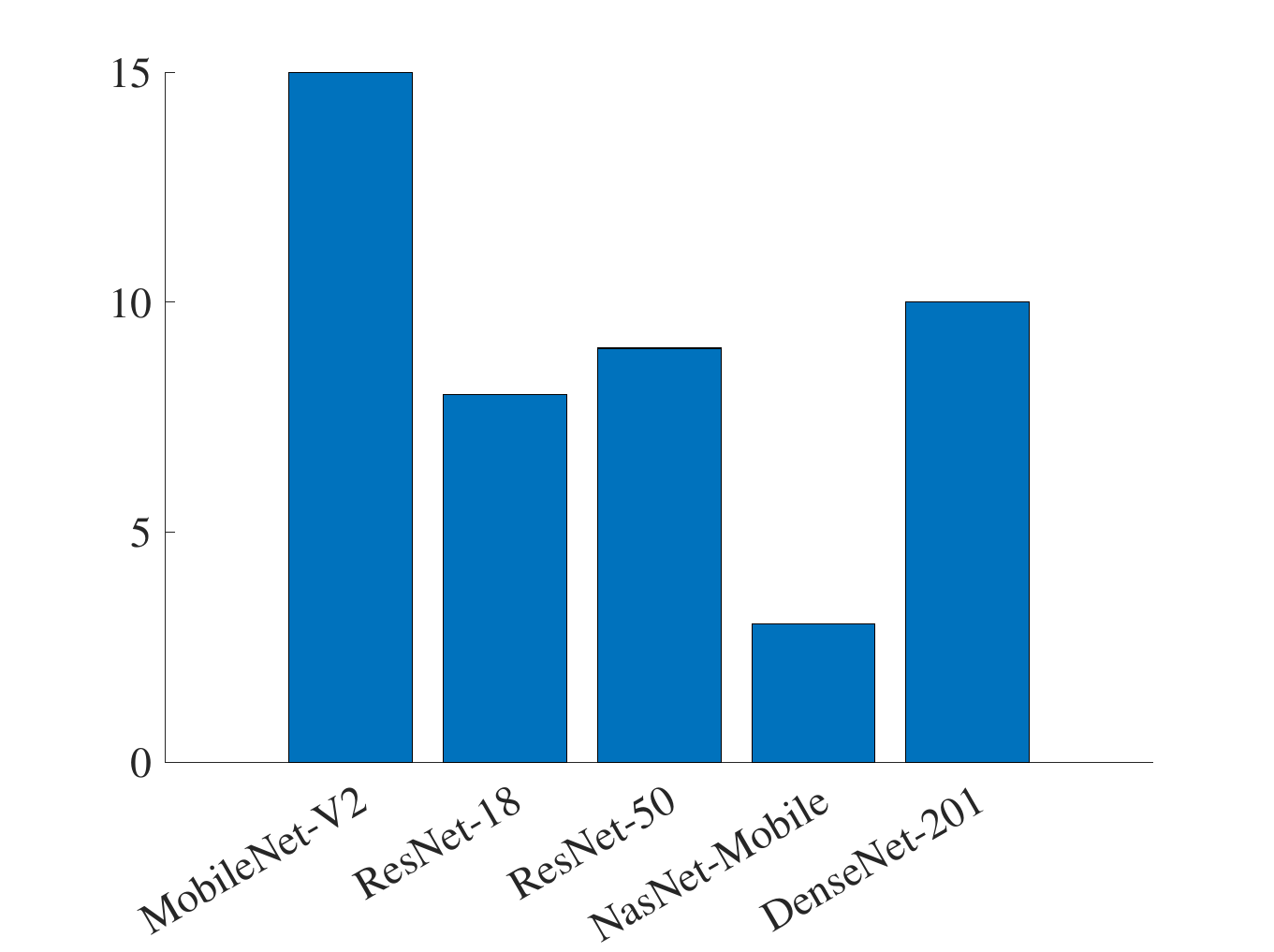}\label{fig:rankingLCA16} } 
        \subfigure[Ranking of LCA set with batch size 64.]{\includegraphics[scale=0.255]{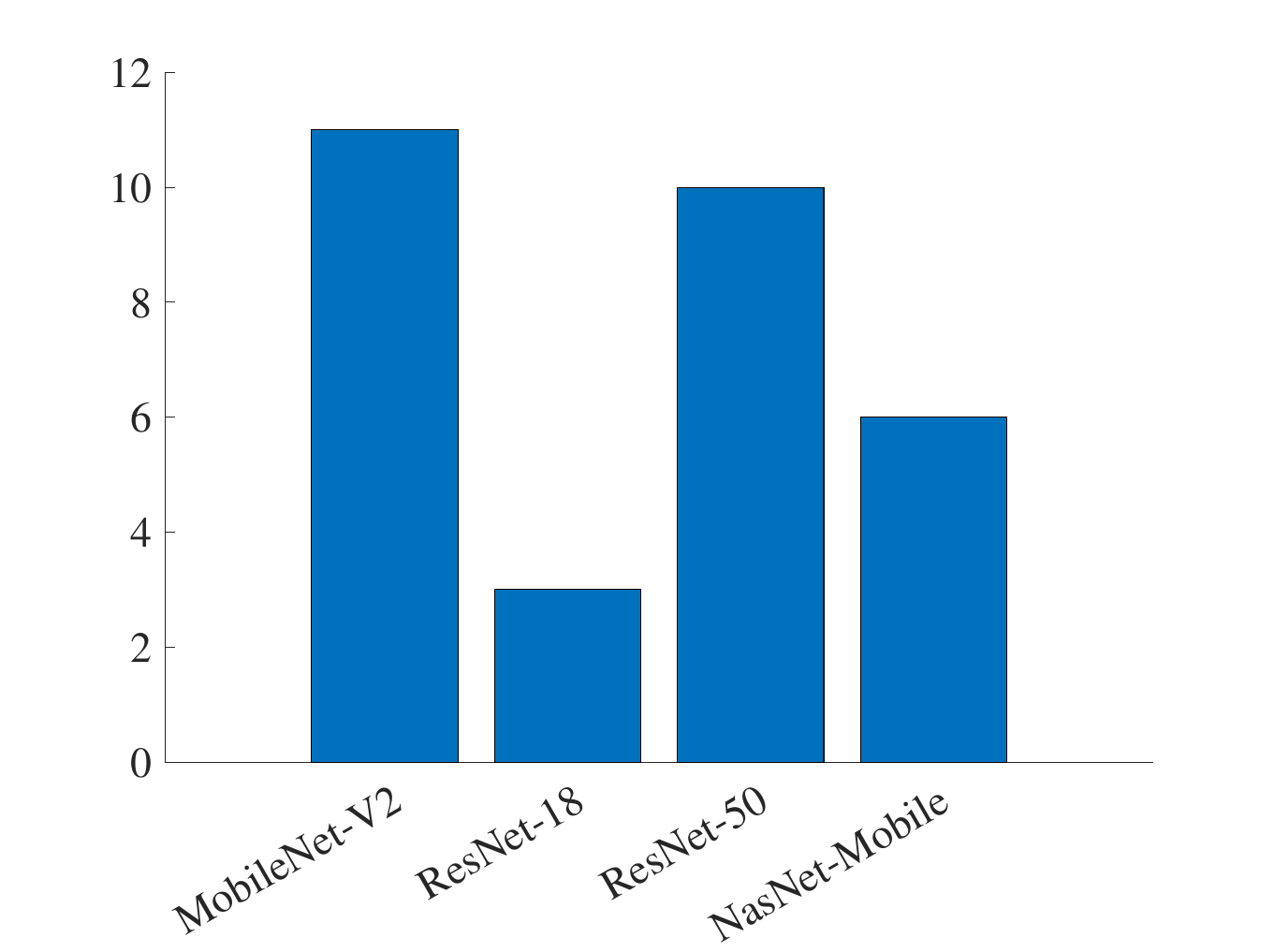} \label{fig:rankingLCA64} }
        
	\subfigure[Ranking of RCA set with batch size 16.]{\includegraphics[scale=0.255]{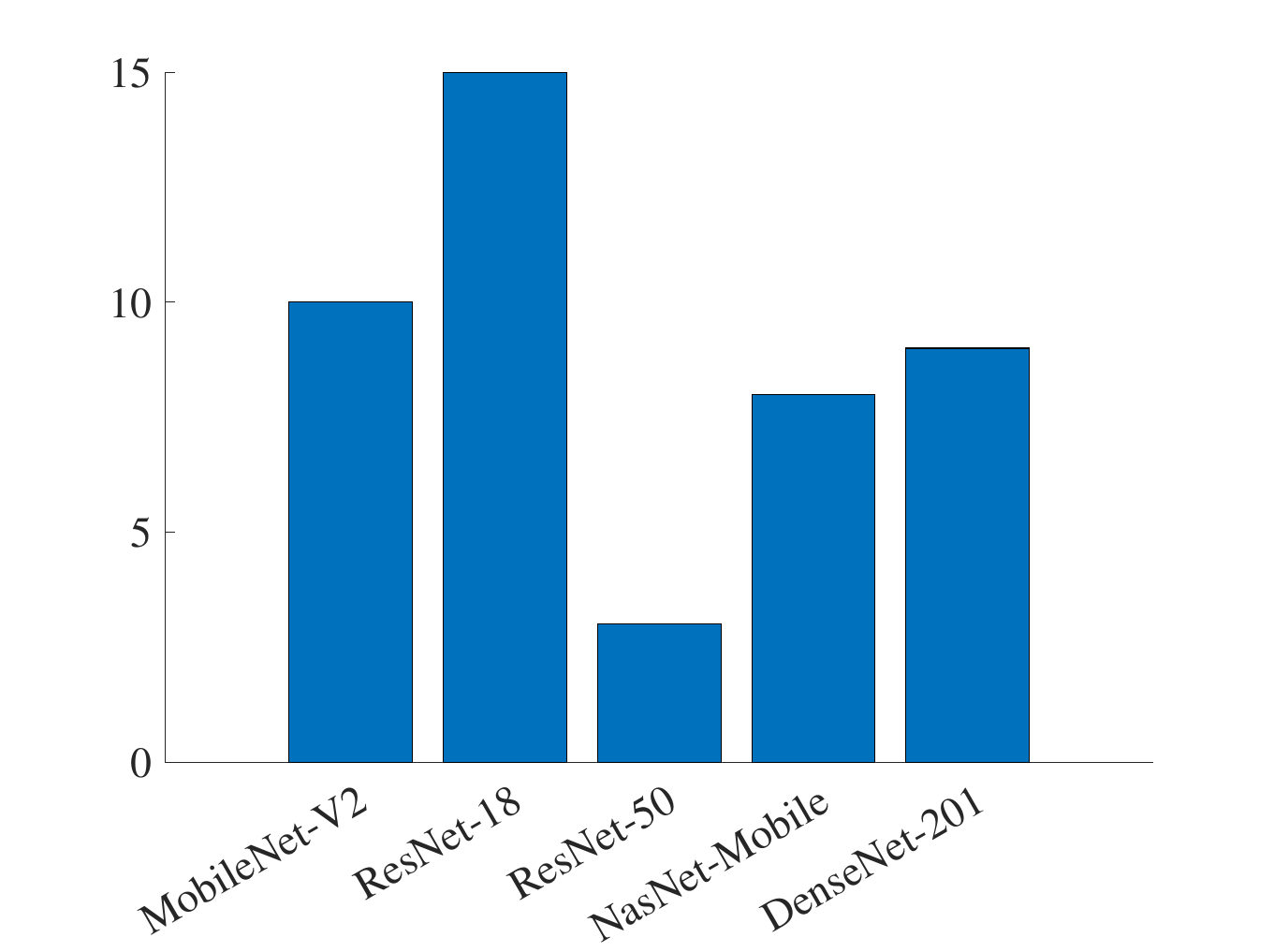} \label{fig:rankingRCA16}}  
	\subfigure[Ranking of RCA set with batch size 64.]{\includegraphics[scale=0.255]{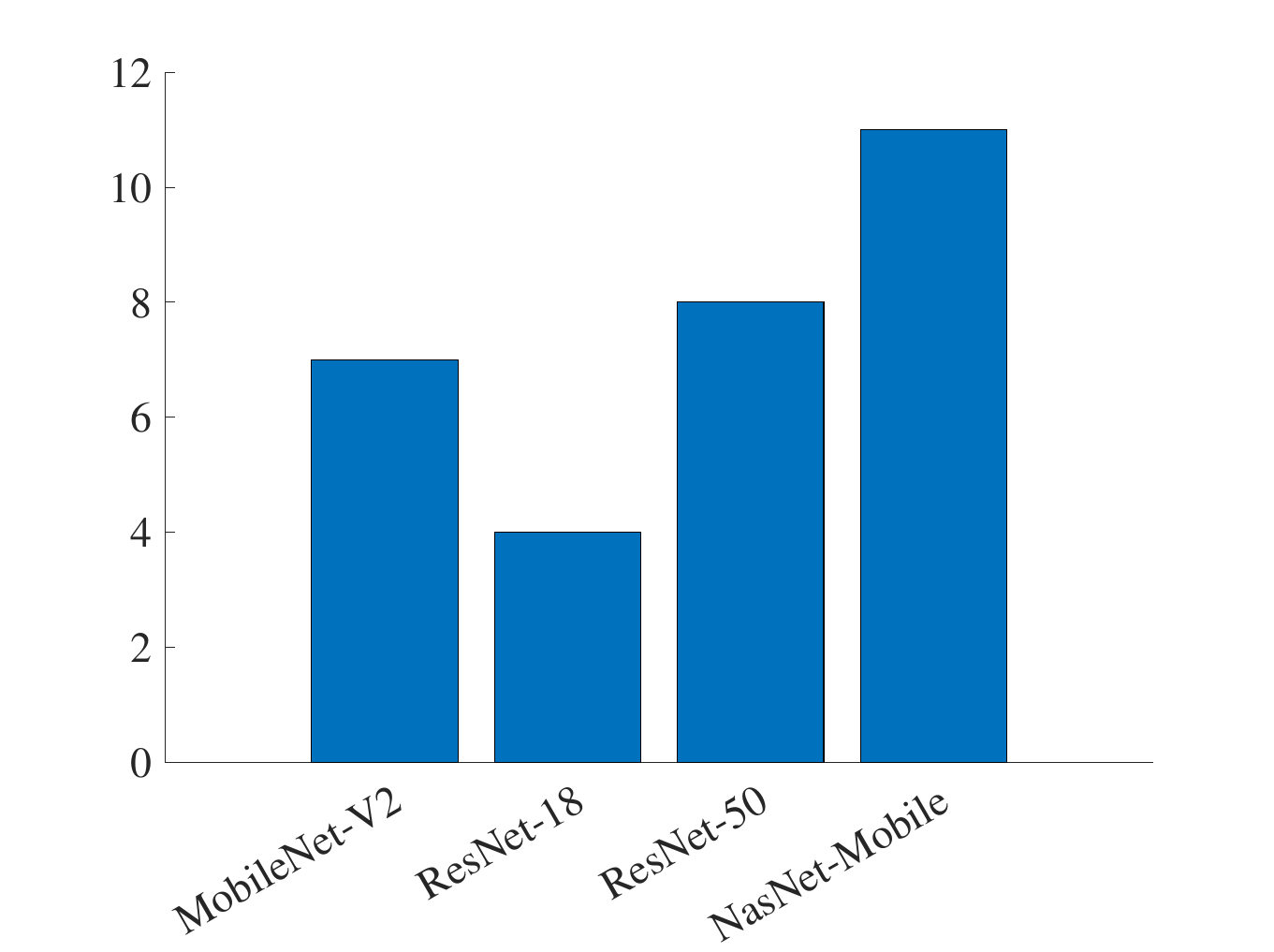} \label{fig:rankingRCA64} }  
 
	\caption{Ranking of methods considering F-measure, Balanced Accuracy and Accuracy obtained for the LCA and RCA subsets using batch sizes 16 and 64.}
	\label{fig:ranking}
\end{figure*}

To study the best suitable model for these inputs, a ranking was implemented to evaluate the results attained considering the three measures. The rankings obtained are reported in Fig. \ref{fig:ranking}, scoring the methods by set and batch size. The scores were calculated by sorting the obtained values of a measure in ascending order since a higher value is better according to the considered measures, meaning that the highest value will be in the last position. The position indicates the obtained scores. There are five and four methods for batch sizes 16 and 64, respectively. Therefore, the maximum possible score is 15 points and 12, respectively, indicating that the method attained the highest values in the three measures.

Focusing on the LCA set, Fig. \ref{fig:rankingLCA16} and \ref{fig:rankingLCA64}, the best outcomes are produced by MobileNet-V2, obtaining 15 points and 11 points with a batch size of 16 and 64, respectively.  Although comparing the results obtained in Table \ref{tab:10fold}, the best result is produced with a batch size of 64, which reached the maximum balanced accuracy, 0.673, and the second best F-measure and Accuracy, 0.810 and 0.732, respectively.

However, in the case of RCA set, Fig. \ref{fig:rankingRCA16} and \ref{fig:rankingRCA64}, ResNet-18 and NasNet-Mobile are the architectures that obtain higher performance, being NasNet-Mobile with a batch size of 64 which returns the best Balanced Accuracy and Accuracy values, 0.658 and 0.744, respectively, and the second best F-measure, 0.826, according to the attained results reported in Table \ref{tab:10fold}.



\begin{table*}[ht]
    \centering
    \caption{Balanced Accuracy, F-measure, and Accuracy results obtained on the test set for LCA and RCA images using 10-fold stratified cross-validation, five convolutional network architectures, and different batch size. The highest values by columns are shown in bold.}
    \resizebox{\textwidth}{!}{
    \begin{tabular}{|l|c|ccc|ccc|}
    \hline
     \multirow{2}{*}{Model} & \multirow{2}{*}{\shortstack{Batch \\ Size}} & \multicolumn{3}{c|}{LCA} & \multicolumn{3}{c|}{RCA} \bigstrut \\
    \cline{3-8} 
        &  & Balanced Accuracy & F-measure & Accuracy & Balanced Accuracy & F-measure & Accuracy \\ 
        \hline
        \multirow{2}{*}{\emph{MobileNet-V2}} 
         & 16 & $0.668 \pm 0.014$ & $0.805 \pm 0.013$ & $0.725 \pm 0.015$ & $0.648 \pm 0.023$ & $0.811 \pm 0.018$ & $0.726 \pm 0.023$  \\ 
         & 64 & $\boldsymbol{0.673 \pm 0.026}$ & $0.810 \pm 0.020$ & $\boldsymbol{0.732 \pm 0.025}$ & $0.641 \pm 0.041$ & $0.806 \pm 0.026$ & $0.719 \pm 0.034$ \\ 
         \hline

        \multirow{2}{*}{\emph{ResNet-18}} 
         & 16 & $0.642 \pm 0.026$ & $0.796 \pm 0.024$ & $0.710 \pm 0.029$ & $0.658 \pm 0.043$ & $0.825 \pm 0.023$ & $0.743 \pm 0.031$ \\ 
         & 64 & $0.632 \pm 0.013$ & $0.779 \pm 0.009$ & $0.691 \pm 0.010$ & $0.624 \pm 0.045$ & $0.792 \pm 0.031$ & $0.702 \pm 0.039$ \\ 
         \hline

        \multirow{2}{*}{\emph{ResNet-50}}
         & 16 & $0.664 \pm 0.035$ & $0.793 \pm 0.029$ & $0.713 \pm 0.036$ & $0.618 \pm 0.022$ & $0.799 \pm 0.009$ & $0.705 \pm 0.013$ \\ 
         & 64 & $0.645 \pm 0.044$ & $\boldsymbol{0.814 \pm 0.021}$ & $0.728 \pm 0.029$ & $0.620 \pm 0.029$ & $\boldsymbol{0.830 \pm 0.025}$ & $0.738 \pm 0.031$ \\ 
         \hline

        \multirow{2}{*}{\emph{NasNet-Mobile}}
         & 16 & $0.634 \pm 0.047$ & $0.789 \pm 0.023$ & $0.701 \pm 0.033$ & $0.651 \pm 0.054$ & $0.804 \pm 0.058$ & $0.723 \pm 0.067$ \\ 
         & 64 & $0.637 \pm 0.043$ & $0.783 \pm 0.020$ & $0.696 \pm 0.029$ & $\boldsymbol{0.658 \pm 0.034}$ & $0.826 \pm 0.033$ & $\boldsymbol{0.744 \pm 0.039}$ \\ 
         \hline
        
        \multirow{1}{*}{\emph{DenseNet-201}}
         & 16 & $0.641 \pm 0.024$ & $0.802 \pm 0.022$ & $0.715 \pm 0.027$ & $0.633 \pm 0.037$ & $0.812 \pm 0.029$ & $0.723 \pm 0.037$ \\ 
        \hline
        
    \end{tabular}
    } \label{tab:10fold}
\end{table*}

\section{Discussion}
\label{sec:discussion}

In this section, some important aspects to be considered of our proposal are discussed.

\begin{itemize}
    \item In this work, all degrees of lesions are considered, while other studies only include severe lesions \cite{au2018automated, shu2021deep} or exclude those images in which more than one lesion appears \cite{au2018automated}. Nevertheless, our results indicate a fair-to-high performance, which means that the dataset is functional, but also that this classification task is complex and challenging.
    \item Despite the fact that a Balanced Accuracy between 0.65 and 0.67 was obtained, it represents a good performance since this measure depends on specificity and recall, which quantifies an average of how correctly both classes are classified, and the ``non-lesion'' class is worse classified than the ``lesion'' class. This is due to the fact that there are fewer ``non-lesion'' images than ``lesion'' images, although data augmentation was applied to equalize both classes. Since the ``lesion'' class includes all degrees of lesion, non-severe lesions can be incorrectly classified as “non-lesion”.
    \item Although the DenseNet model usually has a good performance \cite{guendel2018learning,thurnhofer2021convolutional,chauhan2021optimization}, we could not evaluate it with a batch size of 64 due to memory requirements. Nevertheless, training this model with a batch size of 16 yielded good results in terms of F-measure. Therefore, better results will be expected by training with higher batch sizes.
    \item Finally, for LCA images, the selected configurations outcomes similar results, where the MobileNet-V2 was the most suitable model independently of the batch size. However, the results reported a higher variability for RCA images, since the most suitable models depend on the batch size. Thus, ResNet-18 is the best model for batch size 16, whereas for batch size 64 it is the worst model. This could be because the selected configurations were chosen with 5fold cross-validation results for the LCA images, and these configurations were applied to classify the RCA images.
\end{itemize}

\section{Conclusions}
\label{sec:conclusions}
The coronary dataset published in this work aims to provide the research community with a conscientious and exhaustive scientific resource. It can serve as a benchmark for both algorithm implementations and medical staff to train their abilities on angiographic assessment of CAD severity. Considering the researchers’ feedback, a set of utilities and the already extensive dataset will be regularly revised and expanded.

Experiments were designed to try the functionality of the dataset, which was divided into LCA and RCA images. Five well-known classification architectures were trained and tested using augmented data to get an overview of the performance classification of the ``lesion'' and ``non-lesion'' images. Experiments showed that the most suitable models to solve this problem were MobileNet-V2 for LCA images and NasNet-Mobile for RCA images, getting fair-to-high outcomes, around 80\% F-measure and Accuracy, and 65\% Balanced Accuracy. These results were obtained considering a wide range of lesion levels and support the idea that this classification task is complex, setting up a challenge for physicians and computer-aided diagnosis systems. Therefore, the provided dataset gives the scientific community a starting point to improve CAD detection.

Given the complexity of the classification problem posed, other architectures could be tested in order to keep evaluating the performance of different models. Also, future works will focus on classifying severe lesions images, as well as trying to identify each type of severity. The use of image patches to detect and classify the arteries present in this region will help to improve classification rates.

\section*{Declarations}
\subsection*{Data availability}
CADICA dataset is open-access available at the Mendeley Data repository with the data identification number: 10.17632/p9bpx9ctcv.1, and direct URL to data: \url{https://data.mendeley.com/datasets/p9bpx9ctcv/1}.

\subsection*{Author Contibutions}
All authors listed have made a substantial, direct, and intellectual
contribution to the work, and approved it for publication. 

\subsection*{Declaration of competing interest}
The authors declare that the research was conducted in the absence
of any commercial or financial relationships that could be construed as a potential conflict of interest.

\subsection*{Conflicts of Interest}
The authors declare that they have no conflicts of interest to report regarding the present study

\subsection*{Acknowledgment}
This work is partially supported by the Autonomous Government of Andalusia (Spain) under project UMA20-FEDERJA-108, project name Detection, characterization and prognosis value of the non-obstructive coronary disease with deep learning, and also by the Ministry of Science and Innovation of Spain, grant number PID2022-136764OA-I00, project name Automated Detection of Non Lesional Focal Epilepsy by Probabilistic Diffusion Deep Neural Models. It includes funds from the European Regional Development Fund (ERDF). It is also partially supported by the University of M\'alaga (Spain) under grants B1-2019\_01, project name Anomaly detection on roads by moving cameras; B1-2019\_02, project name Self-Organizing Neural Systems for Non-Stationary Environments; B1-2021\_20, project name Detection of coronary stenosis using deep learning applied to coronary angiography; B4-2022, project name Intelligent Clinical Decision Support System for Non-Obstructive Coronary Artery Disease in Coronarographies; B1-2022\_14, project name Detecci\'on de trayectorias an\'omalas de veh\'iculos en c\'amaras de tr\'afico; and by the Fundaci\'on Unicaja under project PUNI-003\_2023, project name Intelligent System to Help the Clinical Diagnosis of Non-Obstructive Coronary Artery Disease in Coronary Angiography. The authors thankfully acknowledge the computer resources, technical expertise and assistance provided by the SCBI (Supercomputing and Bioinformatics) center of the University of M\'alaga. They also gratefully acknowledge the support of NVIDIA Corporation with the donation of a RTX A6000 GPU with 48Gb. The authors also thankfully acknowledge the grant of the Universidad de M\'alaga and the Instituto de Investigaci\'on Biom\'edica de M\'alaga y Plataforma en Nanomedicina-IBIMA Plataforma BIONAND.

\printbibliography

\vspace{3em}

\end{document}